\documentclass[aps,twocolumn,showpacs,amsmath,amssymb]{revtex4}
\usepackage{amsthm}
\usepackage{latexsym}
\usepackage{amsfonts}
\usepackage{bbm,dsfont}
\usepackage{color}
\usepackage{graphicx}

\newtheorem{proposition?}{Proposition?}
\newtheorem{theorem}{Theorem}
\newtheorem{lemma}{Lemma}
\newtheorem{corollary}{Corollary}

\newtheorem{definition}{Definition}

\newcommand{\real}{\mathbb R} 
\renewcommand{\natural}{\mathbb N} 
\newcommand{\half}{\tfrac{1}{2}} 

\newcommand{\hi}{\mathcal{H}} 
\newcommand{\hik}{\mathcal{K}} 
\newcommand{\lh}{\mathcal{L(H)}} 
\newcommand{\lk}{\mathcal{L(K)}} 
\newcommand{\ip}[2]{\left\langle\,#1\,|\,#2\,\right\rangle} 
\newcommand{\kb}[2]{|#1\rangle\langle#2|} 
\newcommand{\no}[1]{\left\|#1\right\|} 
\newcommand{\tr}[1]{\textrm{tr}\left[#1\right]} 
\newcommand{\id}{\mathbf{1}} 
\newcommand{\nul}{0} 

\newcommand{\obs}{\mathfrak{O}}
\newcommand{\A}{\mathsf{A}}
\newcommand{\B}{\mathsf{B}}
\newcommand{\C}{\mathsf{C}}
\newcommand{\Y}{\mathsf{Y}}
\newcommand{\R}{\mathsf{R}}

\newcommand{\I}{\mathcal{I}}

\newcommand{\chan}{\mathfrak{C}}
\newcommand{\ch}[1]{\mathfrak{C}_{#1}}

\newcommand{\pleq}{\preceq}

  \makeatletter
  \def\mathcomposite{%
     \@ifstar
        {\def\@mathcomposite@option{%
            \baselineskip\z@skip\lineskiplimit-\maxdimen}%
         \@mathcomposite}%
        {\let\@mathcomposite@option\offinterlineskip
         \@mathcomposite}}
  \def\@mathcomposite{%
     \@ifnextchar[\@@mathcomposite{\@@mathcomposite[0]}}
  \def\@@mathcomposite[#1]#2#3#4{%
     #2{\mathchoice
        {\@mathcomposite@{#1}{#3}{#4}\displaystyle{1}}%
        {\@mathcomposite@{#1}{#3}{#4}\textstyle{1}}%
        {\@mathcomposite@{#1}{#3}{#4}%
         \scriptstyle\defaultscriptratio}%
        {\@mathcomposite@{#1}{#3}{#4}%
         \scriptscriptstyle\defaultscriptscriptratio}}}
  \def\@mathcomposite@#1#2#3#4#5{%
     \vcenter{\m@th\@mathcomposite@option
        \dimen@\f@size\p@\dimen@#1\dimen@\dimen@#5\dimen@
        \divide\dimen@ 18
        \edef\@mathcomposite@skipamount{\the\dimen@}%
        \ialign{\hfil$#4##$\hfil\cr
           #2\crcr
           \noalign{\vskip\@mathcomposite@skipamount}%
           #3\crcr}}}
  \makeatother

\newcommand{\psleq}{\precsim}

\begin{document}

\title[Qualitative Noise-Disturbance Relation]{Qualitative Noise-Disturbance Relation for Quantum Measurements}

\author{Teiko Heinosaari}
\email{teiko.heinosaari@utu.fi}
\affiliation{Turku Centre for Quantum Physics, Department of Physics and Astronomy, University of Turku, Finland}

\author{Takayuki Miyadera}
\email{miyadera@nucleng.kyoto-u.ac.jp}
\affiliation{Department of Nuclear Engineering, Kyoto University - 6068501
Kyoto, Japan}


\begin{abstract}The inherent connection between noise and disturbance is one of the most fundamental features of quantum measurements. 
In the two well-known extreme cases a measurement either makes no disturbance but then has to be totally noisy or is as accurate as possible but then has to disturb so much that all subsequent measurements become redundant.
Most of the measurements are, however, something between these two extremes.
We derive a structural connection between certain order relations defined on observables and channels, and we explain how this connection properly explains the trade-off between noise and disturbance.
A link to a quantitative noise-disturbance relation is demonstrated.
\end{abstract}

\maketitle

\section{Introduction}\label{sec:intro}

The inherent connection between noise and disturbance is one of the most fundamental features of quantum measurements. 
On the one hand, a measurement cannot give any information without disturbing the object system.
On the other hand, a noisier (less informative) measurement can be implemented with less disturbance than a sharper measurement.
Roughly speaking, more noise means that measurement outcome distributions become broader, while disturbance is reflected in the measurement outcome statistics of subsequent measurements.
In the most extreme case, the disturbance inherent in a measurement makes all subsequent measurements useless as far as the original input state is concerned.

Various trade-off inequalities between noise (or information) and disturbance are known, all depending on different quantification of these notions, see e.g. \cite{Ozawa03pra,Banaszek06,MiIm06pra, Maccone07,BuHaHo08prl,KrScWe08}.
All these trade-off inequalities are revealing different aspects of the interplay between noise and disturbance in quantum measurements.
In this work we present a relation between certain important forms of noise and disturbance which is qualitative in nature and not based on any specific quantifications of noise and disturbance.
Our result is a structural connection between observables and channels. 
More precisely, we show that a certain partial order in the set of equivalence classes of quantum observables (positive operator valued measures) corresponds to an inclusion of the related subsets of quantum channels (trace preserving completely positive maps).
As we will explain, this correspondence has a clear interpretation as a noise-disturbance relationship since it shows how the possible state transformations are limited to more noisy ones if the measurement is required to be more accurate.
Due to its simplicity and generality, we believe that our qualitative noise-disturbance relation can be seen as a common origin of many quantitative noise-disturbance inequalities.

To give a preliminary idea on the coming developments, we recall two well-known special situations. (See e.g. \cite{Ozawa01pra,HeWo10} for general results that cover these cases.)
First, let us consider a measurement in an orthonormal basis $\{\varphi_j \}_{j=1}^d$.
If $\varrho$ is an input state, then the measurement outcome probabilities are $\ip{\varphi_j}{\varrho \varphi_j}$.
The output state is a mixture $\sum_j \ip{\varphi_j}{\varrho \varphi_j} \xi_j$, where $\xi_1,\xi_2,\ldots$ are states that depend on the measurement device but not on the input state.
Hence, a measurement in an orthonormal basis is sharp but disturbs a lot.
A completely different kind of measurement is such that we do nothing on the input state but we just throw a dice to produce measurement outcome probabilities. 
This measurement has maximum amount of noise, but it can be implemented without disturbing the input state at all.

Most of measurements belong to the intermediate area between the two previously described extreme cases.
Namely, they contain some additional noise and can be measured in a way that implies some disturbance. 
More noise should allow for a less disturbing measurement, and vice versa.
It is exactly this kind of intuitive trade-off that we will turn into an exact theorem. 

In the rest of the paper $\hi$ is a fixed Hilbert space related to the input system. 
The dimension of $\hi$ can be either finite or countably infinite.
We denote by $\lh$ the set of all bounded operators on $\hi$. 
A quantum measurement produces measurement outcomes and conditional output states.
The mapping from input states to measurement outcome statistics is called an observable, while the mapping from input states to unconditional output states (i.e. average over conditional output states) is called a channel \cite{MLQT12}.
We will briefly recall some of the basic properties of observables and channels before proving our main results, Theorem \ref{th:ideal} and Theorem \ref{thm:main}.

\section{Order structure of observables}

A quantum observable with finite or countably infinite number of outcomes is described by a mapping $x\mapsto \A(x)$ such that each $\A(x)\in \lh$ is a positive operator (i.e. $\ip{\psi}{\A(x)\psi}\geq 0$ for all $\psi\in\hi$) and $\sum_x \A(x)=\id$, where $\id$ is the identity operator on $\hi$.
The labeling of measurement outcomes is not important for the questions that we will investigate, hence we assume that the outcome set of all our observables is $\natural=\{1,2,\ldots\}$. 
We denote by $\obs$ the set of all observables on $\hi$.
Let us remark that it is possible that $\A(x)=\nul$ for some outcomes $x$, hence e.g. observables with only a finite number of outcomes are included in $\obs$ by adding zero operators.
For each observable $\A$, we denote by $\Omega_\A \subseteq \natural$ the set of all outcomes $x$ with $\A(x)\neq 0$. 

\begin{figure}
\begin{center}
\includegraphics[width=8.0cm]{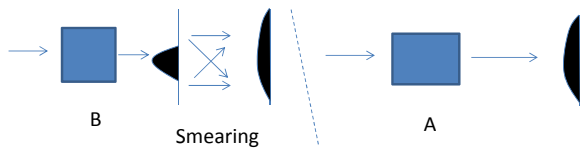}
\end{center}
\caption{\label{fig:smearing} If $\A\pleq\B$, then a measurement of $\A$ can be simulated by a measurement of $\B$ and a classical channel $M$ applied to the measurement outcome distribution.}
\end{figure}

By a \emph{stochastic matrix} we mean a real matrix $[M_{xy}]$, $x,y\in\natural$ such that $M_{xy}\geq 0$ and $\sum_x M_{xy}=1$.
Given two observables $\A$ and $\B$, we denote $\A\pleq\B$ if there exists a stochastic matrix $M$ such that
\begin{equation}
\A(x)=\sum_y M_{xy} \B(y) 
\end{equation}
for all $x\in\natural$.
 The relation $\pleq$ is a preordering in $\obs$, i.e., $\A\pleq\A$ for every observable $\A$, and if $\A\pleq\B$ and $\B\pleq\C$, then $\A\pleq \C$.
This preordering structure has been called with different names in the literature; non-ideality \cite{MaMu90a}, smearing \cite{OQP97}, post-processing \cite{BuDaKePeWe05}.
The physical meaning of the relation is that if $\A\pleq\B$, then 
(in the level of measurement outcome statistics) a measurement of $\A$ can be simulated by a measurement of $\B$ and a classical channel applied to the measurement outcome distribution; see Fig. \ref{fig:smearing}.
In this sense, $\B$ is superior to $\A$.
The physical mechanism of the additional noise of $\A$ compared to $\B$ is typically related to a weaker measurement coupling or impurities in the ancilla state. 
We refer to \cite{OQP97} for some realistic examples. 

Let us note that it is possible to have $\A\pleq\B$ and $\B\pleq\A$ even if $\A\neq\B$ \cite{Heinonen05}. 
For this reason, it is often appropriate to study equivalence classes of observables rather than single observables.
We denote $\A\simeq \B$ if and only if both $\A\pleq \B$ and $\B\pleq \A$ hold. 
Then $\simeq$ is an equivalence relation and the equivalence class of $\A$ is denoted by $[\A]$. 
Physically speaking, the equivalence class $[\A]$ contains all observables $\B$ that are like $\A$ in all relevant ways but may differ by the ordering of measurement outcomes or some other irrelevant detail.
We introduce the set of equivalence classes $\obs^\sim := \obs / \simeq$ and the preorder $\pleq$ then induces a partial order
 $\pleq$ on $\obs^\sim$  by $[\A] \pleq [\B]$ if and only if $\A \pleq \B$. 
(We use the same symbol $\pleq$ for these two different relations, but this should not cause a confusion.)
It is easy to see that in the partially ordered set $\obs^\sim$, there exists the least element but there is no greatest element. 
Namely, an observable $\C$ defined by $\C(1)=\id$, $\C(j)=\nul$ for $j\neq 1$ is a representative of the least element since
for every $\A\in \obs$, the equality $\id =\sum_x \A(x)$ holds. 
The equivalence class $[\C]$ consists of all 'coin tossing observables', i.e.,
\begin{equation*}
[\C]=\{ \C_p| \C_p(x)=p(x)\id, 0\leq p(x)\leq 1, \sum_x p(x)=1\} \, .
\end{equation*}
The measurement outcome of an observable $\C_p$ is determined by a fixed probability distribution $p$ and does not depend on the input state at all.  

To see that there is no greatest element in $\obs^\sim$, suppose on the contrary that $\B$ is such.
Let $\{\varphi_x\}$ be an orthonormal basis and define an observable $\A$ by $\A(x) =\kb{\varphi_x}{\varphi_x}$. 
Then the condition $\kb{\varphi_x}{\varphi_x}=\sum_y M_{xy} \B(y)$ implies that 
every $\B(y)$ is proportional to some $\kb{\varphi_x}{\varphi_x}$. 
But since this should hold for arbitrary orthonormal basis $\{\varphi_x\}$, we must have $\B(y)=0$. 
This contradicts the fact that $\sum_y \B(y)=\id$.

\section{Order structure of channels}

A measurement process yields a probability distribution of measurement outcomes, but it also causes a change 
of the input state. 
This state transformation is described by a quantum channel. 
In the Schr\"odinger picture a channel is a completely positive map that maps an input state to an output state. 
We allow the output state to belong to a different operator space $\lk$ than the input state.
For instance, a mapping $\varrho\mapsto \varrho \otimes \xi$, where $\xi\in\lk$ is a fixed state, is a valid channel.
This particular channel adds an ancilla system in a state $\xi$ to the original system.

For the purposes of this paper, it is more convenient to use the Heisenberg picture description for channels.
In the Heisenberg picture a channel is defined as a normal completely positive map $\Lambda: \lk \to \lh$ satisfying $\Lambda(\id_{\hik})=\id_{\hi}$, where $\hik$ is the output Hilbert space. 
The Schr\"odinger picture description $\Lambda^S$ of a channel $\Lambda$ can be obtained from the relation
\begin{equation}
\tr{\Lambda^S(\varrho)C}=\tr{\varrho \Lambda(C)} \, , 
\end{equation}
true for all states $\varrho\in\lh$ and operators $C\in\lk$.

We denote by $\chan$ the set of all channels from an arbitrary output space $\lk$ to the fixed input space $\lh$. For two channels $\Lambda_1, \Lambda_2\in \chan
$, 
we denote $\Lambda_1 \psleq \Lambda_2$ if there exists a channel $\mathcal{E}$ such that $\Lambda_1 =\Lambda_2 \circ \mathcal{E}$. 
This relation is analogous to the one defined for observables, and the physical meaning of $\Lambda_1 \psleq \Lambda_2$ is that $\Lambda_1$ can be simulated by using $\Lambda_2$ and $\mathcal{E}$ sequentially.
It is easy to see that this relation is a preorder but not a partial order. 

As in the case of observables, it is often convenient to work on the level of equivalence classes of channels.
If $\Lambda_1 \psleq \Lambda_2$ and $\Lambda_2 
\psleq \Lambda_1$ hold, then we denote  
$\Lambda_1 \sim \Lambda_2$. 
The relation $\sim$ is an equivalence relation, which 
allows us to introduce the set of equivalence classes
$\chan^\sim:=\chan / \sim$. 
The equivalence class of a channel $\Lambda$ is denoted by $[\Lambda] \in \chan^\sim$, and a natural partial order 
$\psleq$ is 
introduced by 
$[\Lambda_1] \psleq [\Lambda_2]$ if and only if $\Lambda_1 \psleq \Lambda_2$. 

In the partially order set $\chan^\sim$, there exists the greatest element and the least element. 
Namely, for a state $\varrho\in\lh$, we define 
\begin{equation}\label{eq:greatest-channel}
\Lambda_{\varrho}: \lh \to \lh \, , \quad \Lambda_{\varrho}(C)=\tr{\varrho C}\id_{\hi} \, .
\end{equation} 
Then for any $\Lambda: \lk \to \lh$, the equation $\Lambda_{\varrho}=\Lambda \circ \Lambda'_{\varrho}$ holds, where $\Lambda'_{\varrho}: \lh \to \lk$ is defined as $\Lambda'_{\varrho}(C)=\tr{\varrho C} \id_{\hik}$.
 Thus $[\Lambda_{\varrho}]$ is the least element in $\chan^\sim$. 
On the other hand, the identity channel $id:\lh \to \lh$ defined by $id(C)= C$ 
for all $C\in \lh$ belongs to the greatest equivalence class since any channel $\Lambda$ satisfies $\Lambda= id \circ \Lambda$.

\section{Compatible observables and channels}

A unifying description of the measurement outcome statistics and the state change under a measurement process is given by the notion of an instrument \cite{QTOS76}. 
In the Schr\"odinger picture an instrument is a mapping $(x,\varrho)\mapsto \I^S_x(\varrho)$ such that $\tr{\I^S_x(\varrho)}$ is the probability of obtaining an outcome $x$ and the operator $\widetilde{\varrho}_x=\I^S_x(\varrho) / \tr{\I^S_x(\varrho)}$ is the conditional output state under the condition that a measurement outcome $x$ is obtained. 
The unconditional output state is thus given by $\widetilde{\varrho} \equiv \sum_x \I^S_x(\varrho)$. 
The map $\varrho \mapsto \widetilde{\varrho}$ is a channel in the Schr\"odinger picture. 
We recall that every instrument has a measurement model consisting of an ancillary system and its initial state, a measurement interaction and a pointer observable on the ancillary system \cite{Ozawa84}. 
As in the case of channels, the Heisenberg picture for instruments is convenient for our purposes. 
An instrument in the Heisenberg picture 
is defined by a family of normal completely positive maps 
$\I_x : \lk \to \lh$ whose sum $\sum_x \I_x$ is a channel. 

We are interested in what pairs of observables and channels can belong to the same measurement process.
Therefore, the following concept is useful.

\begin{definition}\label{def:a-channel}
Let $\A$ be an observable on $\hi$.
A channel $\Lambda:\lk\to\lh$ is an $\A$-channel if there exists an instrument $\I$ such that
\begin{equation*}
\I_x(\id_\hik)=\A(x) \, , \qquad \sum_x \I_x(C) = \Lambda(C) \, .
\end{equation*}
We denote by $\ch{\A}$ the set of all $\A$-channels.
\end{definition}

In other words, $\Lambda$ is an $\A$-channel if $\Lambda$ and $\A$ are parts of a single instrument $\I$. 
Following \cite{HeMiRe12}, we call such devices $\Lambda$ and $\A$ \emph{compatible}.

Let $\A$ be an observable on $\lh$. 
If $\Lambda \in \chan
$ is an $\A$-channel, 
any $\Lambda' \in \chan$ satisfying $\Lambda' \psleq \Lambda$ is also an $\A$-channel. 
Namely, suppose there exists an instrument $\I$ such that $\Lambda =\sum_x \I_x$ and $\I_x(\id)=\A(x)$.
If $\Lambda'=\Lambda\circ\mathcal{E}$ for some channel $\mathcal{E}$, then we have $\Lambda'=\sum_x \I_x \circ \mathcal{E}$ and 
$(\I_x \circ \mathcal{E})(\id)=\A(x)$. 
Consequently, if $\Lambda$ is an $\A$-channel, 
any $\Lambda' \in [\Lambda]$ is also an $\A$-channel. 
Thus, a subset $\ch{\A}^\sim$ of $\chan^\sim$ is naturally introduced
as $\ch{\A}^\sim=\{[\Lambda]|\ \Lambda \mbox{ is an }\A \mbox{-channel}\}$.  
It is easy to see that the partially ordered set $\ch{\A}^\sim$ contains the least element.
Namely, $\ch{\A}^\sim$ contains the least element of $\chan^\sim$, the equivalence class $[\Lambda_{\varrho}]$, introduced in \eqref{eq:greatest-channel}.
The fact that $\Lambda_{\varrho}$ belongs to $\ch{\A}$ for any observable $\A$ relates to the possibility of performing a destructive measurement; we can always measure $\A$, destroy the system and prepare a state $\varrho$. 

A less obvious and more interesting fact is that the partially ordered set $\ch{\A}^\sim$ contains the greatest element.
To construct a channel belonging to the greatest element of $\ch{\A}^\sim$, let $(\hik, \hat{\A}, K)$ be a Naimark dilation of $\A$; $\hik$ is a Hilbert space, $K: \hi \to \hik$ is an isometry, and $\hat{\A}$ is a projection-valued measure (PVM)
 on $\hik$ satisfying 
$K^* \hat{\A}(x) K =\A(x)$ for all $x\in\natural$. 
We define a channel $\Lambda_{\A}: \lk \to \lh$ by
\begin{eqnarray}\label{eq:LambdaA} 
\Lambda_{\A} (C) =\sum_x K^* \hat{\A}(x) C \hat{\A}(x) K \, .
 \end{eqnarray}
To see that $\Lambda_{\A}$ is an $\A$-channel, we define an instrument $\I$ by
\begin{equation}
\I_x(C)=K^* \hat{\A}(x) C \hat{\A}(x) K \, .
\end{equation}
Then $\sum_x \I_x = \Lambda_{\A}$ and $\I_x(\id)=K^*\hat{\A}(x) K=\A(x)$.
Although the construction of $\Lambda_{\A}$ relies on 
the choice of the Naimark dilation $(\hik, \hat{\A}, K)$, the following arguments 
do not depend on this choice. 
From now on, we will always assume that a Naimark dilation $(\hik, \hat{\A}, K)$ has been fixed  for each observable $\A$, hence also $\Lambda_{\A}$ is defined for each $\A$.
 
\begin{theorem}\label{th:ideal}
Let $\A$ be an observable. 
The set $\chan_{\A}$ of all $\A$-channels consists of all channels that are below $\Lambda_\A$, i.e.,
\begin{equation}
\chan_{\A}=\{ \Lambda \in \chan \, |\,  \Lambda \psleq \Lambda_\A \} \, .
\end{equation}
Thus, $\ch{\A}^\sim$ has the greatest element $[\Lambda_{\A}]$ and
\begin{equation}
\chan^\sim_{\A}=\{ [\Lambda] \in \chan^\sim \, |\,  [\Lambda] \psleq [\Lambda_\A]\} \, .
\end{equation}
\end{theorem}

\begin{figure}
\begin{center}
\includegraphics[width=7.0cm]{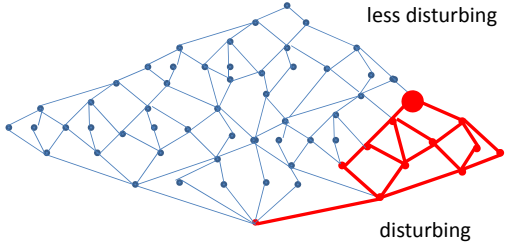}
\end{center}
\caption{\label{fig:ideal} The set of $\chan^\sim$ of all equivalence classes of channels is here illustrated as a net of points.
A downward path between two points means that the lower equivalence class is below the upper one in the partial order $\precsim$. The set $\chan^\sim_{\A}$ (red) consists of all elements that are below a single element $[\Lambda_\A]$ (big dot).} 
\end{figure}

The result of Theorem \ref{th:ideal} is illustrated in Fig. \ref{fig:ideal}.
From the mathematical point of view, the set $\chan^\sim_{\A}$ generated by a single element $[\Lambda_\A]$ is called a \emph{principal ideal}, which is the minimal ideal containing $[\Lambda_\A]$. 

From the physical point of view, Theorem \ref{th:ideal} tells that there is a specific channel $\Lambda_\A$ among all $\A$-channels, and all other $\A$-channels can be obtained from $\Lambda_\A$ by applying a suitable channel after the measurement. It is even justified to call $\Lambda_\A$ a \emph{least disturbing} $\A$-channel since an additional channel after it cannot decrease the caused disturbance.

\begin{proof}[\textbf{Proof of Theorem \ref{th:ideal}}]
We have already seen that $\chan_{\A} \supseteq \{ \Lambda \in \chan \, |\, \Lambda \psleq \Lambda_\A \}$, hence we need to show that the inclusion holds in the other direction as well.

Let $\Lambda: \mathcal{L}(\hik') \to \lh$ be an $\A$-channel.
To prove that $\Lambda \psleq \Lambda_\A$, we first fix a minimal Stinespring dilation $(\hik'', V)$ of $\Lambda$.
Thus, $\hik''$ is a Hilbert space, $V: \hi \to \hik' \otimes \hik''$ is an isometry
satisfying $\Lambda (C) =V^* (C\otimes \id)V$ and the set $(\mathcal{L}(\hik') \otimes \id)V\hi$ is dense in $\hik' \otimes \hik''$.
Since $\Lambda$ is an $\A$-channel, we can apply the Radon-Nikodym theorem of CP-maps \cite{Arveson69, Raginsky03} to conclude that
there exists a unique observable $\R$ on $\mathcal{L}(\hik'')$ satisfying
\begin{eqnarray*}
\A(x)=V^* (\id \otimes \R(x))V
\end{eqnarray*}
for all $x\in\natural$.
For each $x\in\Omega_{\A}$, we define an operator $c_x: \hi \to \hik'\otimes \hik''$ by
$
c_x :=(\id \otimes \R(x)^{1/2})V
$.
Then for any $C\in \mathcal{L}(\hik')$, we have
\begin{equation}\label{eqC}
\Lambda(C) = \sum_x c_x^* (C\otimes \id )c_x.
\end{equation}
Since $c_x$ satisfies $c_x^* c_x =\A(x)$,
by the polar decomposition theorem there exists an isometry $W_x : \hi \to \hik' \otimes \hik''$
satisfying
\begin{eqnarray}\label{cxWx}
c_x=W_x\sqrt{\A(x)} \, , 
\end{eqnarray}
and therefore
\begin{equation}\label{eqC-2}
\Lambda(C) = \sum_x \sqrt{\A(x)} W_x^\ast  (C\otimes \id ) W_x\sqrt{\A(x)} \, .
\end{equation}
We note that if $\dim\hi=\infty$, then the polar decomposition theorem states that $W_x$ is a partial isometry (and not necessarily isometry).
However, in our setting it is possible to extend the partial isometry to an isometric operator.
This additional argument is given in the Appendix.

Let $(\hik, \hat{\A}, K)$ be the Naimark dilation of $\A$.
The relationship 
$K^* \hat{\A}(x) K =\A(x)$ implies that 
there exists an isometry $J_x : \hi \to \hik$ satisfying 
\begin{equation}\label{eq:Jx}
\hat{\A}(x) K = J_x \sqrt{\A(x)} \, .
\end{equation}
Again, the argument why $J_x$ is an isometry and not just a partial isometry is given in the Appendix.
Inserting \eqref{eq:Jx} into \eqref{eqC-2} gives 
\begin{eqnarray*}
\Lambda(C) =\sum_x 
K^* \hat{\A}(x) J_x W_x^* (C\otimes \id) W_x J^*_x \hat{\A}(x) K. 
\end{eqnarray*} 

Finally, fix an arbitrary state $\rho$ on $\hik'$. 
We define 
\begin{eqnarray*}
\mathcal{E}(C)&:=&
\sum_x \hat{\A}(x) J_x W^*_x(C\otimes \id_{\hik'})W_x J^*_x \hat{\A}(x)
\nonumber \\
&&
+ \mbox{tr}[\rho C] (\id -\sum_x \hat{\A}(x) J_x J^*_x \hat{\A}(x)  ).
\end{eqnarray*} 
Then $\mathcal{E}$ is a channel and  
\begin{eqnarray*}
&&
\Lambda_{\A}\circ \mathcal{E} (C) =\Lambda(C) + \\ 
&+& \mbox{tr}[\rho C] 
\left( 
\sum_x K^* \hat{\A}(x) K 
- \sum_x K^* \hat{\A}(x) J_x J_x^* \hat{\A}(x) K
\right)
\\
&=&
\Lambda(C) 
+\mbox{tr}[\rho C] 
\left (\id - 
\sum_x \sqrt{\A(x)}\sqrt{\A(x)} \right) 
=\Lambda(C). 
\end{eqnarray*}
Thus we obtain $\Lambda = \Lambda_{\A} \circ \mathcal{E}$, implying that  $\Lambda \psleq \Lambda_\A$.
\end{proof}

Let us emphasize that the existence of a least disturbing channel is generally guaranteed only if the output space $\hik$ is not fixed. 
This is a noteworthy difference to the analogous result on instruments.
In that case, a least disturbing instrument (in the sense of conditional post processing) exists even if we fix $\hik=\hi$;  
see e.g. Theorem 7.2 in \cite{Hayashi}.

\section{Noise -- Disturbance Relation}

Suppose that $\A$ and $\B$ are two observables satisfying $\ch{\B}\subseteq\ch{\A}$.
This means that every $\B$-channel is also $\A$-channel, so even without any quantification of noise we can conclude that it is possible to measure $\A$ with less or equal disturbance than generated in any measurement of $\B$.
In other words, the unavoidable disturbance related to $\A$ is smaller than or equal to the unavoidable disturbance related to $\B$.
This qualitative description of disturbance will be the basis of the forthcoming noise - disturbance relation. 

The following preliminary observation is easily extracted from our earlier discussion and Theorem \ref{th:ideal}.

\begin{lemma}\label{Lemma:min}
Let $\A$ and $\B$ be two observables. 
Then $\chan_{\B}\subseteq \chan_{\A}$ 
if and only if $\Lambda_{\B} \in \chan_{\A}$. 
\end{lemma}

We are now ready to proceed to our second main result.

\begin{theorem}\label{thm:main}(Qualitative noise-disturbance relation)
Let $\A$ and $\B$ be two observables.
Then $\A\pleq\B$ if and only if $\ch{\B}\subseteq\ch{\A}$.
\end{theorem}

\begin{figure}
\begin{center}
\includegraphics[width=8.0cm]{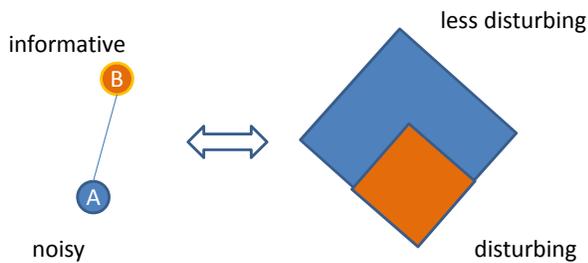}
\end{center}
\caption{\label{fig:main} Illustration of Theorem \ref{thm:main}:
The smearing relation $\A\pleq\B$ of two observables (left) holds if and only if the associated sets of channels are ordered by inclusion $\ch{\A}\supseteq\ch{\B}$ (right).}
\end{figure}

This result is illustrated in Fig. \ref{fig:main}.
It is already intuitively clear that if an observable $\A$ is noisier than $\B$, then it should be possible to measure $\A$ in a less disturbing way.
The purpose of Theorem \ref{thm:main} is to sharpen and clarify certain aspects of this intuitive idea. 
First of all, Theorem \ref{thm:main} shows that the fundamental trade-off between noise and disturbance is a structural feature of quantum theory that can be expressed even without any quantifications of these notions.  

Perhaps the more surprising part of Theorem  \ref{thm:main} is that the inclusion $\ch{\B}\subseteq\ch{\A}$ implies the smearing relation $\A\pleq\B$. 
In particular, if two observables $\A$ and $\B$ are compatible with exactly the same set of channels, i.e. $\ch{\A}=\ch{\B}$, then $\A$ and $\B$ are equivalent and can thus differ only by some physically irrelevant ways.
Therefore, the set $\ch{\A}$ of all $\A$-channels characterizes the observable $\A$ essentially.
 
In some situations, the smearing relation $\A\pleq\B$ can be seen as too restrictive characterization of noise.
For instance, we may try to use $\A$ as an approximate version of $\B$ even if $\A\pleq\B$ does not hold.
Theorem \ref{thm:main} then implies that the associated sets of channels are not anymore in an inclusion relation. 
This should not be understood in the sense that the smearing relation $\A\pleq\B$ is the only reasonable way to characterize noise, but that it determines the setting where the related disturbances are indisputably ordered, no matter on the quantification.
A consideration on some more specific class of measurements may well justify another kind of comparison of observables and channels.

\begin{proof}[\textbf{Proof of Theorem \ref{thm:main}}.]
The \emph{only if}-part: Suppose that $\A\pleq\B$, hence there exists a stochastic matrix $M$ such that
$\A(x)=\sum_y M_{xy} \B(y)$.
Let $\Lambda:\lk\to\lh$ be a $\B$-channel, meaning that there exists an instrument $\I$ such that
\begin{equation*}
\I_y(\id_\hik)=\B(y) \, , \qquad \sum_y \I_y(C) = \Lambda(C) \, .
\end{equation*}
We define an instrument $\I'$ by the formula
$\I'_x:= \sum_{y} M_{xy} \I_y $. 
Then it is easy to see 
$\sum_x \I'_x=\Lambda$
and
$\I'_x(\id_{\hik})=\A(x)$. 
Therefore, $\Lambda$ is an $\A$-channel.
Since $\Lambda$ was an arbitrary $\B$-channel, we conclude that $\ch{\B}\subseteq\ch{\A}$.

The \emph{if}-part: 
By Lemma \ref{Lemma:min} we have $\Lambda_{\B} \in \chan_{\A}$. 
A Stinespring representation of $\Lambda_{\B}$ is given by 
an isometry $V: \hi \to \hik \otimes \hik'$,
\begin{eqnarray*}
V\psi = \sum_{x\in \Omega_{\B}} \hat{\B}(x) K \psi \otimes e_x \, ,  
\end{eqnarray*}
where $\hik'$ is a Hilbert space with the dimension  equal to the cardinality of $\Omega_{\B}$ and $\{e_x\}$ is an orthonormal basis of $\hik'$. 
Since $\Lambda_{\B}$ is compatible with $\A$, 
then it follows from the Radon-Nikodym theorem of CP-maps \cite{Arveson69, Raginsky03} that there exists an observable $\Y$ acting on $\hik'$ such that  
\begin{eqnarray*}
\A(y)=V^*(\id \otimes \Y(y))V 
\end{eqnarray*}
for all $y\in\natural$.
(In case the Stinespring representation is not minimal, 
the uniqueness of $\Y$ drops.) 
Thus we obtain for any $\psi \in \hi$, 
\begin{eqnarray*}
\ip{\psi}{\A(y) \psi} 
&=&\sum_{x} \sum_{x'} 
\ip{\hat{B}(x) K \psi}{ \hat{\B}(x') K \psi} \ip{e_x}{ \Y(y) e_{x'}}\\
&=& \ip{\psi}{ \sum_x  \B(x) \ip{e_x}{ \Y(y) e_x} \psi}, 
\end{eqnarray*}
where we used $\hat{B}(x)\hat{B}(x') =\delta_{x x'} \hat{B}(x)$. 
As $M_{yx}: = \ip{e_x}{ \Y(y) e_x}$ is a stochastic matrix, we conclude that $\A\pleq\B$. 
\end{proof}

As a direct consequence of Theorem \ref{th:ideal} and Theorem  \ref{thm:main} we record the following link between the preorderings on observables and channels.
This is, again, one manifestation of the trade-off between noise and disturbance.

\begin{corollary}
Let $\A$ and $\B$ be two observables.
Then $\A\pleq\B$ if and only if their respective least disturbing channels $\Lambda_\A$ and $\Lambda_\B$ satisfy  $\Lambda_\B \psleq \Lambda_\A$.
\end{corollary}

Finally, we note that our results can be applied to any measure of disturbance $D$ on the set of channels that satisfies the natural requirement
$D(\Lambda \circ \mathcal{E}) \geq D(\Lambda)$ for all channels
$\Lambda$ and $\mathcal{E}$. 
Namely, Theorem \ref{th:ideal} implies that any $\A$-channel $\Lambda$ satisfies 
$D(\Lambda)\geq D(\Lambda_{\A})$. 
This enables us to derive a lower bound for the disturbance $D(\Lambda)$ since $\Lambda_{\A}$ has a quite simple form. 
For instance, a very natural disturbance measure $D_{KSW}$ was defined in \cite{KrScWe08} as
\begin{equation*}
D_{KSW}(\Lambda) =
\inf_{\mathcal{R}}\Vert \Lambda \circ \mathcal{R} -id\Vert_{cb} \, , 
\end{equation*}
where the infimum is taken over all channels $\mathcal{R}:\lh\to\lk$ and $\Vert \cdot \Vert_{cb}$ is the completely bounded norm. 
The function $D_{KSW}$ quantifies the quality of the best available decoding channel $\mathcal{R}$ for $\Lambda$, 
and is easily shown to satisfy $D_{KSW}(\Lambda \circ \mathcal{E})\geq D_{KSW}(\Lambda)$. 

It was proved in \cite{KrScWe08} that  $D_{KSW}(\Lambda)$ is bounded by the distance between conjugate channel and completely depolarizing channels. 
By using this result, we can show the following.

\begin{theorem}\label{thm:ksw}
Let $\A$ and $\B$ be two observables.
\begin{itemize}

\item[(a)] If $\A\pleq\B$, then there exists an $\A$-channel $\Lambda_0$ that can be decoded with better or equal quality than any $\B$-channel in the sense that $D_{KSW}(\Lambda)\geq D_{KSW}(\Lambda_0)$ for all $\B$-channels $\Lambda$.
 
\item[(b)] Every $\A$-channel $\Lambda$ satisfies  
\begin{align}\label{eq:ksw}
D_{KSW}(\Lambda) \geq \frac{1}{16}\sup_{x\in\Omega_\A} \bigl( \no{\A(x)} +\no{ \id -\A(x)} -1 \bigr)^2, 
\end{align}
where $\Vert \cdot \Vert$ is the operator norm on $\lh$.
\end{itemize}
\end{theorem}

The right hand side of \eqref{eq:ksw} is related to one of the functions characterizing sharpness and bias of quantum effects, namely, the quantity $\Vert \A(x)\Vert +\Vert \id -\A(x)\Vert -1$ is the width of the spectrum of $\A(x)$ \cite{Bu09}.
It follows that the right hand side of \eqref{eq:ksw} is zero if and only if $\A$ is a coin tossing observable, expressing the fact that 'no disturbance implies no information'. 

In the other extreme case, the quantity $\Vert \A(x)\Vert +\Vert \id -\A(x)\Vert -1$ takes the maximal value $1$ if and only if the spectrum of $\A(x)$ contains both $0$ and $1$ \cite[Prop. 2]{Bu09}. 
For instance, if $\A$ contains a non-trivial projection $\A(x)$ (i.e. $\A(x)^2=\A(x)$ and $0\neq\A(x)\neq\id$), then Theorem \ref{thm:ksw} gives $D_{KSW}(\Lambda) \geq \frac{1}{16}$ for all $\A$-channels $\Lambda$.
This is a lower bound on the quality of the best available decoding channel for any $\A$-channel.

\begin{proof}[\textbf{Proof of Theorem \ref{thm:ksw}}]
\begin{itemize}
\item[(a)] We choose $\Lambda_0=\Lambda_\A$ and then the claim is a direct consequence of Theorems \ref{th:ideal} and \ref{thm:main}.

\item[(b)] Let $\Lambda$ be a channel compatible with $\A$.
As was explained above, we have
\begin{equation}\label{eq:ksw-smaller}
D_{KSW}(\Lambda) \geq D_{KSW}(\Lambda_{\A}) \, .
\end{equation} 
Thus, in the following we estimate $D_{KSW}(\Lambda_{\A})$ and this will lead to a lower bound for $D_{KSW}(\Lambda)$.
The channel $\Lambda_{\A}$ has a Stinespring representation 
$(\mathcal{K}', V)$, where $\mathcal{K}' =\mathbf{C}^{|\Omega_{\A}|}$ 
($|\Omega_{\A}|$ may be infinity) and 
$V$ is defined by 
\begin{eqnarray*}
V\psi=\sum_x \hat{\A}(x) K \psi \otimes e_x \, , 
\end{eqnarray*}
where $\{e_x \}$ is an orthonormal basis of $\mathcal{K}'$. 
Its conjugate channel $\Lambda^c: \mathcal{L}(\mathcal{K}')\to \lh$ is  
\begin{eqnarray*}
\Lambda^c(C) =\sum_x \ip{ e_x}{C e_x} \A(x) \, .
\end{eqnarray*}

Let us denote the completely depolarizing channel with respect to a state $\sigma$ on $\mathcal{K}'$ by 
$S_{\sigma}$, i.e., $S_{\sigma}(C)=\mbox{tr}[\sigma C]\id$. 
According to \cite[Thm. 3]{KrScWe08}, there exists 
$\sigma$ satisfying 
\begin{eqnarray*}
\Vert \Lambda^c - S_{\sigma}\Vert_{cb} 
\leq 2 D(\Lambda_{\A})^{1/2} \, .
\end{eqnarray*}
Thus we have to estimate 
$\inf_{\sigma}\Vert \Lambda^c -S_{\sigma}\Vert_{cb}$. 
Let us denote by $\Vert \cdot \Vert_{\infty}$ the operator norm of channels.
As we have
\begin{eqnarray*}
\inf_{\sigma}\Vert\Lambda^c-S_{\sigma}\Vert_{cb}
&\geq& \inf_{\sigma}\Vert \Lambda^c-S_{\sigma}\Vert_{\infty} \\
&\geq&  \inf_{\sigma}\sup_{E: projection}\Vert \Lambda^c(E) -S_{\sigma}(E)
\Vert, 
\end{eqnarray*}
it holds that for each $x$, 
\begin{eqnarray*}
\inf_{\sigma}\Vert \Lambda^{c} -S_{\sigma}\Vert_{cb} 
&\geq& \inf_{\sigma} \Vert \Lambda^c(|e_x\rangle \langle e_x| )
-S_{\sigma}(|e_x \rangle \langle e_x|)\Vert \\
&=&\inf_{\sigma}\Vert \A(x) - \ip{e_x}{\sigma e_x}\id\Vert
\\
&=& \inf_{0\leq p\leq 1} \Vert \A(x) -p \id\Vert
\\
&=& \frac{\Vert \A(x)\Vert +\Vert \id -\A(x)\Vert -1}{2}.
\end{eqnarray*} 
(For the last equality, see e.g. \cite{Bu09}.)
We have thus proved that
\begin{align}\label{eq:ksw-almost}
\frac{1}{4} \bigl( \no{\A(x)} +\no{ \id -\A(x)} -1 \bigr) \leq  D(\Lambda_{\A})^{1/2} 
\end{align}
for each $x\in\natural$.
From \eqref{eq:ksw-smaller} and \eqref{eq:ksw-almost} follows \eqref{eq:ksw}.
\end{itemize} 
\end{proof}

\section{Example: Binary qubit measurements}

The simplest kind of measurements are binary (i.e. two-outcome) measurements on a qubit system.
For each vector $\vec{v}\in\real^3$ with $\no{\vec{v}}\leq 1$, we define a binary qubit observable $\A^{\vec{v}}$ by
$\A^{\vec{v}}(\pm 1)=\half ( \id  \pm \vec{v}\cdot\vec{\sigma} )
. $
It is easy to see that $\A^{\vec{w}} \pleq \A^{\vec{v}}$ if and only if $\vec{w}$ and $\vec{v}$ are parallel vectors and $\no{\vec{w}}\leq\no{\vec{v}}$.
To demonstrate how this order structure of observables is reflected in the measurement disturbance, let us consider the L\"uders measurements for the above type of qubit observables.
The L\"uders instrument related to $\A^{\vec{v}}$ is defined as
$
\I^{\vec{v}}_x(C)=\sqrt{\A^{\vec{v}}(x)}C \sqrt{\A^{\vec{v}}(x)} \, , \quad x=\pm 1
.$
The corresponding channel is
$
\Lambda^{\vec{v}} = \I^{\vec{v}}_{1} + \I^{\vec{v}}_{-1} = \lambda\ id + (1-\lambda) \ \mathcal{V}
$,  
where 
\begin{equation}\label{eq:qubit-unitary}
 \mathcal{V}(C)=1/\no{\vec{v}}^2 \ \vec{v}\cdot\vec{\sigma} C \vec{v}\cdot\vec{\sigma} \, , \quad
\lambda = \frac{ 1+ \sqrt{1-\no{\vec{v}}^2}}{2} \, .
\end{equation}
Let us note that the unitary channel $\mathcal{V}$ depends on the direction of $\vec{v}$ but not on its norm,
while the weight $\lambda$ depends on the norm of $\vec{v}$ but not on its direction.
Applying Theorem \ref{thm:main} for two observables $\A^{\vec{v}}$ and $\A^{\vec{w}}$ with parallel vectors $\vec{v}$ and $\vec{w}$, we conclude that for two parameters $\lambda,\mu\in[\half,1]$ and a unitary channel $\mathcal{V}$ defined in \eqref{eq:qubit-unitary}, there exists a channel $\mathcal{E}$ such that
\begin{align}\label{eq:qubit-circ}
\bigl( \lambda\ id + (1-\lambda) \ \mathcal{V} \bigr) \circ \mathcal{E} =
\bigl( \mu\ id + (1-\mu)\ \mathcal{V} \bigr) \, .
\end{align}
if and only if $\lambda \geq \mu$.
This is in line what we would expect; the sharper the measurement, the smaller must the weight of the identity channel be.
In this example, it is not too difficult to find the concrete form of a channel $\mathcal{E}$ satisfying \eqref{eq:qubit-circ}. 
Namely, for all $\lambda,\lambda'\in[\half,1]$, we obtain
\begin{align}\label{eq:qubit-circ-2}
\bigl( \lambda\ id + (1-\lambda) \ \mathcal{V} \bigr) \circ \bigl( \lambda'\ id + (1-\lambda')\ \mathcal{V} \bigr) =  \nonumber \\
\bigl( (1-\lambda-\lambda'+2\lambda\lambda')\ id + (\lambda+\lambda'-2\lambda\lambda')\ \mathcal{V} \bigr) \, .
\end{align}
Hence, for every $\mu <\lambda$ we can choose $\lambda'=(\mu+\lambda-1)/(2\lambda -1)$ and then \eqref{eq:qubit-circ-2} leads to \eqref{eq:qubit-circ}.

\section{Summary}

Classical and quantum post-processings yield physically meaningful preorderings in the sets of observables and channels, respectively.
When lifted to the sets of equivalence classes, these relations become partial orderings.
The partial orderings can be seen as abstract and general ways to describe certain important forms of noise and disturbance. 
We have proved that the fundamental trade-off between noise and disturbance in quantum measurements takes a very natural form in this framework.
Namely, an observable $\A$ is more noisy than another observable $\B$ if and only if the set of $\A$-channels (the channels that possibly describe the state transformation in some measurement of $\A$) is larger than the set of $\B$-channels.

\section{Appendix: Isometries in the proof of Theorem \ref{th:ideal}}

If $\dim\hi=\infty$, then the polar decomposition theorem states that a bounded operator $C$ can be written as $C=W\sqrt{C^*C}$, where $W$ is a partial isometry.
Generally, $W$ cannot be chosen to be an isometry.
In this Appendix we show that in the two cases treated in Theorem \ref{th:ideal}, partial isometries can be replaced with isometries.

First, we prove that the operator $W_x$ in \eqref{cxWx} can be chosen to be an isometry.
Since $c_x$ satisfies $c_x^* c_x =\A(x)$,
there exists a partial isometry $W^0_x : \hi \to \hik'' \otimes \hik'$
satisfying $c_x = W^0_x \sqrt{\A(x)}$ and $Ker[W^0_x]=Ker[\A(x)]$.
This latter condition implies that $W_x^{0*}W^0_x=P_{Ker[\A(x)]^{\perp}}$ holds, 
where for a subspace $\mathcal{V}\subseteq \hi$ $P_{\mathcal{V}}$ is the projection onto $\mathcal{V}$ and 
$\mathcal{V}^{\perp}$ represents the orthogonal complement of 
$\mathcal{V}$. 
Let us extend $W^0_x$ to an isometry.
We have
$\id-\A(x)=V^*(\id_{\hik'}\otimes (\id_{\hik''}-\R(x)))V_1$.
Thus there exists a uniquely determined
partial isometry $W'_x$ satisfying
\begin{equation*}
(\id_{\hik''}\otimes (\id_{\hik_1}-\R(x))^{1/2})V_1=W'_x \sqrt{\id_{\hi}-\A(x)}
\end{equation*}
and $Ker[W'_x]=Ker[\id_{\hi}-\A(x)]$. 
Note that $Ker[\id_{\hi}-\A(x)]^{\perp} \supseteq Ker[\A(x)]$.
Thus we can restrict $W'_x$ to $Ker[\A(x)]$ and write it as
$W^1_x$. It satisfies $W^{1*}_xW^1_x=P_{Ker[\A(x)]}$.
Now it can be shown that $W^{0*}W^{1}=\nul$. 
In fact, we have
\begin{eqnarray*}
c_x^*d_x P_{Ker[\A(x)]} &=& \sqrt{\A(x)}W_x^{0*}W_x^{1} \sqrt{\id_{\hi}-\A(x)}
P_{Ker[\A(x)]}
\\
&=&
\sqrt{\A(x)}W_x^{0*}W_x^{1}.
\end{eqnarray*}
The left-hand side of this equality can be written as 
\begin{eqnarray*}
&&c_x^*d_x P_{Ker[\A(x)]}\\
&=&V^*(\id_{\hik'}\otimes \R(x)^{1/2} 
(\id_{\hik'} -\R(x))^{1/2})V P_{Ker[\A(x)]}
\\
&=&V^*(\id_{\hik'}\otimes (\id_{\hik'} -\R(x))^{1/2}
(\id_{\hik'}\otimes \R(x)^{1/2})V P_{Ker[\A(x)]}.
\end{eqnarray*}
As $(\id_{\hik''}\otimes \R(x)^{1/2})V P_{Ker[\A(x)]} =\nul$ holds, 
we have $\sqrt{\A(x)}W_x^{0*}W_x^{1}=\nul$ and 
$W_x^{0*} W_x^1 =\nul$. 
Thus we can define an isometry $W_x=W^0_x\oplus W^1_x$ on the whole space $\hi$.
Consequently we have obtained
an isometry $W_x: \hi \to \hik'\otimes \hik''$ satisfying $c_x=W_x\sqrt{\A(x)}$. 

Second, we show that the operator $J_x$ in \eqref{eq:Jx} can be chosen to be an isometry.
The relationship 
$K^* \hat{\A}(x) K =\A(x)$ implies that 
there exists a partial isometry $J^0_x : \hi \to \hik$ satisfying 
$\hat{\A}(x) K = J^0_x \sqrt{\A(x)}$ 
and $Ker[J^0_x]=Ker[\A(x)]$. 
Since
\begin{equation}
K^*(\id-\hat{\A}(x))K=\id-\A(x)
\end{equation}
holds, there exists a partial isometry $J'_x: \hi \to \hik$ satisfying 
\begin{equation}
\id-\hat{\A}(x)=J'_x \sqrt{\id-\A(x)}
\end{equation}
and $Ker[J'_x]=Ker[\id-\A(x)]$. 
We denote by $J^1_x$ the restriction of $J'_x$ to $Ker[\A(x)]$. 
Then $J_x:=J^0_x\oplus J^1_x$ is an isometry 
satisfying $\hat{\A}(x)K=J_x\sqrt{\A(x)}$.

\section*{Acknowledgements}

TH acknowledges the financial support from the Academy of Finland (grant no. 138135).
TM thanks JSPS for the financial support (JSPS KAKENHI Grant Numbers 22740078).


\begin{thebibliography}{10}

\bibitem{Ozawa03pra}
M.~Ozawa.
\newblock Universally valid reformulation of the Heisenberg uncertainty
  principle on noise and disturbance in measurement.
\newblock {\em Phys. Rev. A}, 67:042105, 2003.

\bibitem{MiIm06pra}
T.~Miyadera and H.~Imai.
\newblock Information-disturbance theorem for mutually unbiased observables.
\newblock {\em Phys. Rev. A}, 73:042317, 2006.


\bibitem{Banaszek06}
K.~Banaszek.
\newblock Information gain versus state disturbance for a single qubit.
\newblock {\em Open Sys. Information Dyn.}, 13:1--16, 2006.

\bibitem{Maccone07}
L.~Maccone.
\newblock Entropic information-disturbance tradeoff.
\newblock {\em Europhys. Lett.}, 77:40002, 2007.

\bibitem{BuHaHo08prl}
F.~Buscemi, M.~Hayashi, and M.~Horodecki.
\newblock Global information balance in quantum measurements.
\newblock {\em Phys. Rev. Lett.}, 100:210504, 2008.

\bibitem{KrScWe08}
D.~Kretschmann, D.~Schlingemann and R.F.~Werner.
\newblock The Information-Disturbance Tradeoff and the Continuity of Stinespring's Representation.
\newblock {\em IEEE Trans. Inf. Theory}, 54:1708, 2008.

\bibitem{Ozawa01pra}
M.~Ozawa.
\newblock Operations, disturbance, and simultaneous measurability.
\newblock {\em Phys. Rev. A}, 63:032109, 2001.

\bibitem{HeWo10}
T.~Heinosaari and M.M. Wolf.
\newblock Nondisturbing quantum measurements.
\newblock {\em J. Math. Phys.}, 51:092201, 2010.

\bibitem{MLQT12}
T.~Heinosaari and M.~Ziman.
\newblock {\em The Mathematical Language of Quantum Theory}.
\newblock Cambridge University Press, Cambridge, 2012.
\newblock From uncertainty to entanglement.

\bibitem{MaMu90a}
H.~Martens and W.M. {de Muynck}.
\newblock Nonideal quantum measurements.
\newblock {\em Found. Phys.}, 20:255--281, 1990.

\bibitem{OQP97}
P.~Busch, M.~Grabowski, and P.J. Lahti.
\newblock {\em Operational Quantum Physics}.
\newblock Springer-Verlag, Berlin, 1997.
\newblock second corrected printing.

\bibitem{BuDaKePeWe05}
F.~Buscemi, G.M. D'Ariano, M.~Keyl, P.~Perinotti, and R.F. Werner.
\newblock Clean positive operator valued measures.
\newblock {\em J. Math. Phys.}, 46:082109, 2005.

\bibitem{Heinonen05}
T.~Heinonen.
\newblock Optimal measurement in quantum mechanics.
\newblock {\em Phys. Lett. A}, 346:77--86, 2005.

\bibitem{QTOS76}
E.B. Davies.
\newblock {\em Quantum Theory of Open Systems}.
\newblock Academic Press, London, 1976.

\bibitem{Ozawa84}
M.~Ozawa.
\newblock Quantum measuring processes of continuous observables.
\newblock {\em J. Math. Phys.}, 25:79--87, 1984.

\bibitem{HeMiRe12}
T.~Heinosaari, T.~Miyadera, and D.~Reitzner.
\newblock Strongly incompatible quantum devices.
\newblock arXiv:1209.1382 [quant-ph], 2012.

\bibitem{CBMOA03}
V.~Paulsen.
\newblock {\em Completely bounded maps and operator algebras}.
\newblock Cambridge University Press, Cambridge, 2003.

\bibitem{Arveson69}
W.~Arveson.
\newblock Subalgebras of {$C^{\ast} $}-algebras.
\newblock {\em Acta Math.}, 123:141--224, 1969.

\bibitem{Raginsky03}
M.~Raginsky.
\newblock Radon-{N}ikodym derivatives of quantum operations.
\newblock {\em J. Math. Phys.}, 44:5003--5020, 2003.

\bibitem{Hayashi}
M.~Hayashi.
\newblock {\em Quantum Informtion}.
\newblock Springer-Verlag, Berlin, 2006.

\bibitem{Bu09}
P.~Busch, 
\newblock On the sharpness and bias of quantum effects.
\newblock {\em Found. Phys.}, 39:712--730, 2009.

\end{thebibliography}
\end{document}